\documentclass[aps,pre,
twocolumn,
groupedaddress,showpacs]{revtex4-2}

\usepackage{threeparttable}
\usepackage{epsfig}
\usepackage{epstopdf}

\usepackage{graphicx}

\usepackage{amsmath}
\usepackage[T1]{fontenc}
\usepackage{caption}
\usepackage{float} 
\usepackage{dcolumn}
\usepackage{hyperref}
\usepackage{amsthm}
\usepackage{amssymb}

\usepackage{booktabs}
\usepackage{xparse}
\usepackage{tikz}
\usetikzlibrary{calc}

\tikzset{Arrow Style/.style={text=black, font=\boldmath}}

\newcommand{\tikzmark}[1]{%
    \tikz[overlay, remember picture, baseline] \node (#1) {};%
}

\newcommand*{\XShift}{0.5em}
\newcommand*{\YShift}{0.5ex}

\NewDocumentCommand{\DrawArrow}{s O{} m m m}{%
    \begin{tikzpicture}[overlay,remember picture]
        \draw[->, thick, Arrow Style, #2] 
                ($(#3.west)+(\XShift,\YShift)$) -- 
                ($(#4.east)+(-\XShift,\YShift)$)
        node [midway,above] {#5};
    \end{tikzpicture}%
}

\def\d{\mbox{d}}

\begin{document}


\title{Using quantum mechanics for calculation of different infinite sums}
\author{ Petar Mali$^{1}$} \author{Milica Rutonjski $^{1}$} \author{Slobodan Rado\v sevi\' c $^{1}$} \author{Milan Panti\' c $^{1}$} \author{Milica Pavkov-Hrvojevi\' c $^{1}$}
\affiliation{$^1$ Department of Physics, Faculty of Sciences, University of Novi Sad,
Trg Dositeja Obradovi\' ca 4, 21000 Novi Sad, Serbia}

\date{\today}


\begin{abstract}
We demonstrate that certain class of infinite sums can be calculated analytically starting from a specific quantum mechanical problem and using principles of quantum mechanics. For simplicity we illustrate the method by exploring the problem of a particle in a box. Twofold calculation of the mean value of energy for the polynomial wave function inside the well yields even argument $p$ ($p>2$) of Riemann zeta and related functions. This method can be applied to a wide class of exactly solvable quantum mechanical problems which may lead to different infinite sums. Besides, the analysis performed here provides deeper understanding of superposition principle and presents useful exercise for physics students.   
\end{abstract}
\maketitle

\section{Introduction}
\label{intro}
For most physics students, when studying quantum mechanics, the first problem that they encounter is the problem of a particle in a box. Standard quantum mechanics textbooks (see for instance \cite{grifits,sif,shankar}) include it as an early exercise illustrating eigenproblem in infinite-dimensional vector space and superposition principle. Although there are some upgrades to the problem including particle in an infinite potential well with moving walls \cite{zidovi,zidovi2} and the problem of splitting the box into two slightly unequal halves \cite{split}, even standard problem of a particle in a box is rich enough for investigation \cite{ermitovost}. In the problem of a particle in a box, an infinite one-dimensional potential well, the potential is given by 
\begin{equation}
  V(x) =
  \begin{cases}
                                   0, & x \in (0,a) \\
                                   \infty, & x \notin (0,a)
 \end{cases}
\end{equation}
where $a$ is the well width. The eigenstates of Hamiltonian, normalized states with well-defined energy, are given by $\psi_n(x)=\sqrt{\frac{2}{a}}\sin\frac{n \pi x}{a}, n=1,2,3...$ The state of the system $\psi(x)$ is an element of $L^2[0,a]$ -- it is a unit vector which vanishes at $x=0$ and $x=a$. Since eigenstates of Hamiltonian form an orthonormal basis, any state $\psi(x)$ of the particle in the well can be written as a linear combination of the eigenstates of Hamiltonian. The mean (expectation) value of energy in the given state $\psi(x)$ can be calculated by using both 
\begin{equation}
\langle \hat{H} \rangle_{\psi}=\int^a_0 \psi^*(x)\left(-\frac{\hbar^2}{2m}\frac{\d^2}{\d x^2}\right)\psi(x)\d x \label{srednjajedan}, 
\end{equation}
and
\begin{equation}
\langle \hat{H} \rangle_{\psi}=\sum^{\infty}_{n=1}W(E_n)E_n \label{srednjadva}
\end{equation}
where 
\begin{equation}
W(E_n)=|C_n|^2=\Bigg|\int^a_0 \psi_n(x) \psi(x)\d x\Bigg|^2.
\label{verovatnoca}\end{equation}
Hereafter we shall use $\langle \hat{H} \rangle_{\psi} \equiv \langle \hat{H} \rangle$ for brevity. In cases when the state of the system is represented by polynomial $\psi(x)$, there are infinitely many non-zero terms in Eq. (\ref{srednjadva}). The aim of the paper is to use that type
of states for the purpose of evaluating a certain class of infinite sums  appearing in the problem of black body radiation \cite{crnotelo} and thermodynamics
of Fermi \cite{filips,rajh} and Bose \cite{sloba} systems.

\section{Results}
\label{Results}
Taking into account that orthonormal basis $\psi_n(x), n=1,2,3...$ inside the well is of trigonometric type, prepared state $\psi(x)$ needs to be a polynomial inside the well in order to be an infinite superposition of basis vectors. It will be shown that polynomials of different degrees produce different infinite sums for the convenient choice of arguments of Riemann zeta function and closely related infinite sums. The main goal will be to calculate those sums and classify their appearance in relation to the degree of the polynomial $\psi(x)$. We shall start the analysis with the second-degree polynomial since first-degree polynomials cannot satisfy boundary conditions $\psi(0)=\psi(a)=0$.
\subsection{Polynomial wave function of second degree} \label{A}
 The only second-degree polynomial wave function satisfying boundary condition is $\psi(x)=Cx(a-x)$, where $C$ is normalization constant. The wave function is normalized by employing $\int^a_0|\psi(x)|^2\d x=1$. Without loss of generality we choose constant $C$ to be positive, which leads to $\psi(x)=\sqrt{\frac{30}{a^5}}x(a-x)$.

As already mentioned the mean value of energy can be calculated twofold. Using (\ref{srednjajedan}) we obtain
\begin{equation}
\langle \hat{H} \rangle=\int^{a}_0\psi(x)\left(-\frac{\hbar^2}{2m}\frac{\d^2}{\d x^2}\right)\psi(x)\d x=\frac{5\hbar^2}{ma^2}, \label{aav1} 
\end{equation}
whereas the determination of mean value of energy using (\ref{srednjadva}) requires the calculation of probabilities $W(E_n)$ defined by (\ref{verovatnoca}). The integration can be simplified by employing the tabular method of the integration by parts \cite{tabularint} which will be used throughout the paper. The given method for solving integral $\int^a_0 x(a-x)\sin \frac{n \pi x}{a}\d x$ is shown below in detail.  
\[
    \renewcommand{\arraystretch}{1.5}
    \begin{array}{c @{\hspace*{1.0cm}} c}\toprule
       D & I \\\cmidrule{1-2}
      x(a-x)\tikzmark{Left 1} & \tikzmark{Right 1}\sin \frac{n \pi x}{a} \\
      a-2x \tikzmark{Left 2} & \tikzmark{Right 2}-\frac{a}{n \pi}\cos \frac{n \pi x}{a} \\      
      -2  \tikzmark{Left 3} & \tikzmark{Right 3}-(\frac{a}{n \pi})^2 \sin \frac{n \pi x}{a} \\      
      0  \tikzmark{Left 4} & \tikzmark{Right 4}(\frac{a}{n \pi})^3\cos \frac{n \pi x}{a} \\\bottomrule
    \end{array}
\]
\DrawArrow[draw=red]{Left 1}{Right 2}{$+$}%
\DrawArrow[draw=brown]{Left 2}{Right 3}{$-$}%
\DrawArrow[draw=blue]{Left 3}{Right 4}{$+$}%
Here $D$ stands for differentiation of the elements of the first column whereas $I$ denotes integration of the second column elements. The elements connected by arrows are multiplied and added by using alternating signs. Thus, we obtain
\vspace{0.2cm}
\begin{widetext}
\begin{equation}
\int^a_0 x(a-x)\sin \frac{n \pi x}{a}\d x=\Bigg[x(a-x)\left(-\frac{a}{n \pi}\cos \frac{n \pi x}{a}\right)-(a-2x)\left(-\left(\frac{a}{n \pi}\right)^2 \sin \frac{n \pi x}{a}\right)+(-2)\left(\frac{a}{n \pi}\right)^3\cos \frac{n \pi x}{a} \Bigg]\Bigg|^a_0,
\end{equation}
\end{widetext}
where due to the integration boundaries only the third term does not vanish. 
 Hereafter the tabular method will not be explicitly shown, since its application is analogous to the one presented above. The simplification introduced by this method becomes invaluable as the degree of the polynomial grows.
  
The performed integration leads to the following probabilities
\begin{equation}
W(E_n)=\frac{480}{n^6\pi^6}[1-(-1)^n].
\end{equation}
Therefrom, the mean value of energy calculated from (\ref{srednjadva}) reads
\begin{equation}
\langle \hat{H} \rangle=\sum^{\infty}_{n=1}\frac{480}{n^6\pi^6}[1-(-1)^n]\frac{n^2\pi^2\hbar^2}{2ma^2}. \label{sumezadrugistepen}
\end{equation}
The only non-zero terms are obtained for $n=2k+1$. Therefore,
\begin{equation}
\langle \hat{H} \rangle=\frac{960}{\pi^6}\frac{\pi^2\hbar^2}{2ma^2}\sum^{\infty}_{k=0}\frac{1}{(2k+1)^4} \label{aav2}.
\end{equation}
From Eqs. (\ref{aav1}) and (\ref{aav2}) we calculate the following infinite sum
\begin{equation}
\sum^{\infty}_{k=0}\frac{1}{(2k+1)^4}=\frac{\pi^4}{96} \label{nep4}.
\end{equation}
Making use of relation (see \cite{atlas})
\begin{equation}
\sum^{\infty}_{n=1}\frac{(-1)^{n-1}}{n^{p}}=(1-2^{-p+1})\sum^{\infty}_{n=1}\frac{1}{n^{p}} \label{tajnaveza}
\end{equation} 
as well as already used relation
\begin{equation}
\sum^{\infty}_{n=1}\frac{1}{n^p}+\sum^{\infty}_{n=1}\frac{(-1)^{n-1}}{n^p}=2\sum^{\infty}_{n=0}\frac{1}{(2n+1)^p} \label{javnaveza}
\end{equation} for $p=4$,
other two sums $\sum^{\infty}_{n=1}\frac{1}{n^4}$ and $\sum^{\infty}_{n=1}\frac{(-1)^{n-1}}{n^4}$ appearing in (\ref{sumezadrugistepen}) can also be calculated starting from second-degree polynomial only. Calculation leads to $\frac{\pi^4}{90}$ and $\frac{7 \pi^4}{720}$, respectively. It should be noted that sum $\sum^{\infty}_{n=1}\frac{1}{n^p}$ is known as Riemann zeta function $\zeta(p)$,  $\sum^{\infty}_{n=1}\frac{(-1)^{n-1}}{n^p}$ is alternating zeta function, also known as Dirichlet eta function $\eta(p)$, whereas $\sum^{\infty}_{n=0}\frac{1}{(2n+1)^p}$ is Dirichlet lambda function $\lambda(p)$.  Calculation of the mean value of energy in state $\psi(x)=\sqrt{\frac{30}{a^5}}x(a-x)$ is performed in different quantum mechanics books (see for instance \cite{grifits}), however not for purpose of obtaining infinite sums. In particular, sum $\lambda(4)$ is therein taken from math tables.

\subsection{Polynomial wave functions of third degree } \label{B}
In the case of third-degree polynomials there exists an infinite set of wave functions $\psi(x)$ that satisfy boundary conditions. Let us take an arbitrary polynomial wave function of this type 
 $\psi(x)=\sqrt{\frac{210}{a^7}}x(a-x)(a-2x).$ The mean value of energy calculated using (\ref{srednjajedan}) is
\begin{equation}
\langle \hat{H} \rangle=\frac{21\hbar^2}{ma^2}. \label{bav1} 
\end{equation}
On the other hand, following (\ref{srednjadva}) and procedure given in Subsection \ref{A} we obtain
\begin{equation}
\langle \hat{H} \rangle=\sum^{\infty}_{n=1}\frac{30240}{n^6\pi^6}[1+(-1)^n]\frac{n^2\pi^2\hbar^2}{2ma^2}. \label{bav15}
\end{equation}
The only non-zero terms are for $n=2k$:
\begin{equation}
\langle \hat{H} \rangle=\frac{1890 \hbar^2}{ma^2 \pi^4}\sum^{\infty}_{k=1}\frac{1}{k^4} \label{bav2}.
\end{equation}
Hence, from Eqs. (\ref{bav1}) and (\ref{bav2}) we obtain Riemann zeta function $\zeta(4)$
\begin{equation}
\sum^{\infty}_{k=1}\frac{1}{k^4}=\frac{\pi^4}{90}. \label{zetaod4}
\end{equation}
Beside this result, sum $\sum^{\infty}_{n=1}\frac{(-1)^{n-1}}{n^4}$ can also be extracted from Eq. (\ref{bav15}). After having determined $\sum^{\infty}_{n=1}\frac{1}{n^4}$ this sum can be calculated from (\ref{sumezadrugistepen}). However, our classification method will be based on which sums can be obtained starting from the certain degree polynomials. Hence, even without knowing the relation (\ref{tajnaveza}), the system of equations for sums $\sum^{\infty}_{n=1}\frac{1}{n^4}$ and $\sum^{\infty}_{n=1}\frac{(-1)^{n-1}}{n^4}$ can be formed by taking another third-degree polynomial, for instance $\psi(x)=\sqrt{\frac{105}{a^7}}x^2(a-x)$. Calculating the mean value of energy from (\ref{srednjajedan}) and (\ref{srednjadva}) we obtain
\begin{equation}
\langle \hat{H} \rangle=\frac{7\hbar^2}{ma^2} \label{cav1}
\end{equation}
and
\begin{equation}
\langle \hat{H} \rangle=\frac{420 \pi^2 \hbar^2}{\pi^6 ma^2}[5\sum^{\infty}_{n=1}\frac{1}{n^4}-4\sum^{\infty}_{n=1}\frac{(-1)^{n-1}}{n^4}] \label{cav2},
\end{equation}
respectively. 
Using result (\ref{zetaod4})
we get
\begin{equation}
\sum^{\infty}_{n=1}\frac{(-1)^{n-1}}{n^4}=\frac{7\pi^4}{720}.
\end{equation}
We have illustrated the method that was not necessary to employ here, yet we discuss it for methodical reasons since it becomes inevitable for higher-degree polynomials. 
It is now straightforward to calculate sum $\sum^{\infty}_{n=0}\frac{1}{(2n+1)^4}$ using Eq. (\ref{javnaveza}), however we may ask whether it can be obtained without calculating $\sum^{\infty}_{n=1}\frac{1}{n^4}$ and $\sum^{\infty}_{n=1}\frac{(-1)^n}{n^4}$, as in the case of the second-degree polynomial. Namely, any third-degree polynomial wave function inside the well can be written in general form $\psi(x)=x(a-x)Q_1(x)$, where $Q_1(x)$ is a first-degree polynomial. Following that,

\begin{widetext}
\begin{equation}
C_n=\sqrt{\frac{2}{a}}\int^a_0 \psi(x)\sin \frac{n \pi x}{a}\d x=\sqrt{\frac{2}{a}}\left[\left(\frac{a}{n\pi}\right)^3(-1)^n[-2Q_1(a)-2aQ_1'(a)]-\left(\frac{a}{n\pi}\right)^3[-2Q_1(0)+2aQ_1'(0)]\right]. \end{equation}
\end{widetext}
We can see that the only three sums that can be possibly obtained in the case of third-degree polynomial wave functions are $\sum^{\infty}_{n=1}\frac{1}{n^4}$, $\sum^{\infty}_{n=1}\frac{(-1)^{n-1}}{n^4}$ and $\sum^{\infty}_{n=0}\frac{1}{(2n+1)^4}$. The last one appears when
\begin{equation}
-2Q_1(a)-2aQ_1'(a)=-2Q_1(0)+2aQ_1'(0),
\end{equation}
which is equivalent to
\begin{equation}
Q_1(0)-Q_1(a)=2aQ_1'(0). \label{uslov}
\end{equation}
This is however not possible for any first-degree polynomial $Q_1$. Condition (\ref{uslov}) is satisfied only for $Q_1(x)=\mbox{const}$ ($\psi(x)$ is a second-degree polynomial), i.e. there exists no third-degree polynomial wave function from which $\sum^{\infty}_{n=0}\frac{1}{(2n+1)^4}$ could be independently calculated. To understand profoundly why this is the case it is necessary to translate the reference frame so that $x'=x-\frac{a}{2}$. In the new reference frame potential is given by \begin{equation}
  V(x') =
  \begin{cases}
                                   0, & x' \in (-\frac{a}{2},\frac{a}{2}) \\
                                   \infty, & x' \notin (-\frac{a}{2},\frac{a}{2})
 \end{cases}
.\end{equation}
By acting with the translation operator $T(\frac{a}{2})=\mbox{exp}(\frac{a}{2}\frac{d}{d x})$ (see for instance \cite{fermi,grajner}) on the states $\psi_n(x)=\sqrt{\frac{2}{a}}\sin\frac{n\pi x}{a}, n=1,2,3,...$ one can obtain eigenstates in potential $V(x')$ as
\begin{equation}
  \psi_n(x')=\sqrt{\frac{2}{a}}
  \begin{cases}
                                   (-1)^{\frac{n-1}{2}}\cos\frac{n \pi x'}{a}, & \mbox{for odd}\,\, n \\
                                   (-1)^{\frac{n}{2}}\sin \frac{n \pi x'}{a}, & \mbox{for even}\,\, n
 \end{cases}
.\end{equation}
States in the new reference frame are even when $n$ is odd number, and vice versa. By translating the state
$\psi(x)=x(a-x)Q_1(x)=x(a-x)(a_1x+b_1)$ with the same translation operator we get the state $\psi(x')=(x'+\frac{a}{2})(\frac{a}{2}-x')(a_1x'+\frac{a_1a}{2}+b_1)$. The sum $\sum^{\infty}_{n=0}\frac{1}{(2n+1)^4}$ can be solely obtained only in the case when $\psi(-x')=\psi(x')$ and this is possible only when $a_1=0$, which corresponds to the polynomial function of the second degree.

\subsection{Polynomial wave function of degree four and higher} \label{C}
Following the routine from Subsection \ref{B} we choose two arbitrary fourth-degree polynomials, $\sqrt{\frac{252}{a^9}}x^3(a-x)$ and $\sqrt{\frac{630}{a^9}}x^2(a-x)(a-2x)$. The expectation values of energy $\langle \hat{H} \rangle$ in the given states read $\frac{54\hbar^2}{5ma^2}$ and $\frac{24\hbar^2}{ma^2}$, respectively. Analogous procedure yields the system of equations 
\begin{widetext}
\begin{eqnarray}
&& \sum^{\infty}_{n=1} \Big[\frac{36}{n^4 \pi^4}-\frac{288}{n^6\pi^6}[1-(-1)^n]+\frac{1152}{n^8\pi^8}[1-(-1)^n]\Big]=\frac{3}{70} \label{jnamil} \nonumber\\
&&\sum^{\infty}_{n=1}\Big[\frac{4}{n^4\pi^{4}}[17-8(-1)^n]
+ \frac{960}{n^6\pi^6}[(-1)^n-1]+\frac{4608}{n^{8}\pi^{8}}[1-(-1)^n]\Big]=\frac{4}{105}.
\end{eqnarray}
\end{widetext}
Here-from it is obvious that sums $\sum^{\infty}_{n=1}\frac{1}{n^4}$, $\sum^{\infty}_{n=1}\frac{1}{n^6}$, $\sum^{\infty}_{n=1}\frac{1}{n^8}$, $\sum^{\infty}_{n=1}\frac{(-1)^{n-1}}{n^4}$, $\sum^{\infty}_{n=1}\frac{(-1)^{n-1}}{n^6}$, $\sum^{\infty}_{n=1}\frac{(-1)^{n-1}}{n^8}$, $\sum^{\infty}_{n=0}\frac{1}{(2n+1)^4}$, $\sum^{\infty}_{n=0}\frac{1}{(2n+1)^6}$, and $\sum^{\infty}_{n=0}\frac{1}{(2n+1)^8}$ can be obtained starting from the fourth-degree polynomials only. More detailed inspection of (\ref{jnamil}) suggests that one more fourth-degree polynomial has to be used to get the system of equations in closed form, due to the fact that some of the upper-mentioned sums are related (see Eqs. (\ref{tajnaveza}) and (\ref{javnaveza})).

The sum $\sum^{\infty}_{n=0}\frac{1}{(2n+1)^4}$ does not appear in (\ref{jnamil}). However it can be independently calculated in case when at least one of the polynomials $\psi(x)$ is chosen so that $e^{\frac{a}{2}\frac{d}{dx}}\psi(x)$ is an even function, for instance $\psi(x)=\sqrt{\frac{10080}{313a^{9}}}x(a-x)(\frac{a}{2}+x)(\frac{3a}{2}-x)$. It should be noted that three polynomials are still sufficient for obtaining all mentioned sums.

If we proceed to the fifth-degree polynomials, we discover that they do not lead to new sums. This brings us to conclusion that odd-degree polynomials are not as rich in the analytically computable sums as the even-degree ones and therefore are of less interest. In the case of the sixth-degree polynomial wave functions the appropriate choice of polynomials would offer us the possibility to calculate the sums $\sum_{n=1}^{\infty}\frac{1}{n^p}$ and $\sum_{n=1}^{\infty}\frac{(-1)^{n-1}}{n^p}$ as well as $\sum_{n=0}^{\infty}\frac{1}{(2n+1)^p}$ for $p=4,6,8,10,12$. Should we proceed to polynomials of higher degree, the pattern would remain the same. The list of the sums that can be calculated starting from the polynomial wave functions of certain degree, together with the values of those sums, is given in Table \ref{tabelasume}.
\begin{widetext}
\begin{center}
\begin{table}
 
\caption{List of the sums that can be calculated using the $n$th degree polynomial wave functions}
\begin{threeparttable}
 \begin{tabular}{c |c| c |c} 

 \hline\hline
 degree & $\psi(x)$  & $p$ & values of sums \tnote{a}\\ [0.5ex] 
 \hline\hline
 2 & $Cx(a-x)\,\,\,\, \,\qquad \tnote{b} $  & $4$&\vtop{\hbox{\strut \vtop{\hbox{\strut $\zeta(4)=\frac{\pi^4}{90}$}\hbox{\strut $\eta(4)=\frac{7\pi^4}{720}$}}}\hbox{\strut $\lambda(4)= \frac{\pi^4}{96}$}} \\ \hline
 
 3 &\,\, $Cx(a-x)Q_1(x) \,\,\,\, \tnote{c}$ \hspace{1mm}  & $4$  &\vtop{\hbox{\strut \vtop{\hbox{\strut $\zeta(4)=\frac{\pi^4}{90}$}\hbox{\strut $\eta(4)=\frac{7\pi^4}{720}$}}}\hbox{\strut $\lambda(4)= \frac{\pi^4}{96}$}}    \\ \hline
 
 4 & $Cx(a-x)Q_2(x)$  & $4$,$6$,$8$ & \vtop{\hbox{\strut \vtop{\hbox{\strut $\zeta(4)=\frac{\pi^4}{90}$\quad$\zeta(6)=\frac{\pi^6}{945}$\quad $\zeta(8)=\frac{\pi^8}{9450}$}\hbox{\strut  $\eta(4)=\frac{7\pi^4}{720}$\quad $\eta(6)=\frac{31\pi^6}{31240}$\quad $\eta(8)=\frac{127\pi^8}{1209600}$}}  }\hbox{\strut  $\lambda(4)=\frac{\pi^4}{96}$\quad $\lambda(6)=\frac{\pi^6}{960}$\quad $\lambda(8)=\frac{17\pi^8}{161280}$}}   \\ \hline
 
 5 & $Cx(a-x)Q_3(x)$ & $4$,$6$,$8$  & \vtop{\hbox{\strut \vtop{\hbox{\strut $\zeta(4)=\frac{\pi^4}{90}$\quad$\zeta(6)=\frac{\pi^6}{945}$\quad $\zeta(8)=\frac{\pi^8}{9450}$}\hbox{\strut  $\eta(4)=\frac{7\pi^4}{720}$\quad $\eta(6)=\frac{31\pi^6}{31240}$\quad $\eta(8)=\frac{127\pi^8}{1209600}$}}  }\hbox{\strut  $\lambda(4)=\frac{\pi^4}{96}$\quad $\lambda(6)=\frac{\pi^6}{960}$\quad $\lambda(8)=\frac{17\pi^8}{161280}$}}   \\ \hline
 
6 & $Cx(a-x)Q_4(x)$  &$4$,$6$,$8$,$10$,$12$& \vtop{\hbox{\strut \vtop{\hbox{\strut \vtop{\hbox{\strut \vtop{\hbox{\strut \vtop{\hbox{\strut $\zeta(4)=\frac{\pi^4}{90}$\quad$\zeta(6)=\frac{\pi^6}{945}$\quad $\zeta(8)=\frac{\pi^8}{9450}$}\hbox{\strut  $\zeta(10)=\frac{\pi^{10}}{93555}$\quad $\zeta(12)=\frac{691\pi^{12}}{638512875}$}}}\hbox{\strut   $\eta(4)=\frac{7\pi^4}{720}$\quad $\eta(6)=\frac{31\pi^6}{31240}$\quad $\eta(8)=\frac{127\pi^8}{1209600}$}}}\hbox{\strut  $\eta(10)=\frac{73\pi^{10}}{6842880}$\quad $\eta(12)=\frac{1414477\pi^{12}}{1307674368000}$}}}\hbox{\strut  $\lambda(4)=\frac{\pi^4}{96}$\quad $\lambda(6)=\frac{\pi^6}{960}$\quad $\lambda(8)=\frac{17\pi^8}{161280}$}}}\hbox{\strut  $\lambda(10)=\frac{31\pi^{10}}{2903040}$\quad $\lambda(12)=\frac{691\pi^{12}}{638668800}$}} \\                      [1ex] \hline

7 & $Cx(a-x)Q_5(x)$  &$4$,$6$,$8$,$10$,$12$& \vtop{\hbox{\strut \vtop{\hbox{\strut \vtop{\hbox{\strut \vtop{\hbox{\strut \vtop{\hbox{\strut $\zeta(4)=\frac{\pi^4}{90}$\quad$\zeta(6)=\frac{\pi^6}{945}$\quad $\zeta(8)=\frac{\pi^8}{9450}$}\hbox{\strut  $\zeta(10)=\frac{\pi^{10}}{93555}$\quad $\zeta(12)=\frac{691\pi^{12}}{638512875}$}}}\hbox{\strut   $\eta(4)=\frac{7\pi^4}{720}$\quad $\eta(6)=\frac{31\pi^6}{31240}$\quad $\eta(8)=\frac{127\pi^8}{1209600}$}}}\hbox{\strut  $\eta(10)=\frac{73\pi^{10}}{6842880}$\quad $\eta(12)=\frac{1414477\pi^{12}}{1307674368000}$}}}\hbox{\strut  $\lambda(4)=\frac{\pi^4}{96}$\quad $\lambda(6)=\frac{\pi^6}{960}$\quad $\lambda(8)=\frac{17\pi^8}{161280}$}}}\hbox{\strut  $\lambda(10)=\frac{31\pi^{10}}{2903040}$\quad $\lambda(12)=\frac{691\pi^{12}}{638668800}$}} \\                      [1ex] \hline

8 & $Cx(a-x)Q_6(x)$  &$4$,$6$,$8$,$10$,$12$,$14$,$16$& \vtop{\hbox{\strut \vtop{\hbox{\strut \vtop{\hbox{\strut \vtop{\hbox{\strut \vtop{\hbox{\strut \vtop{\hbox{\strut \vtop{\hbox{\strut \vtop{\hbox{\strut $\zeta(4)=\frac{\pi^4}{90}$\quad$\zeta(6)=\frac{\pi^6}{945}$\quad $\zeta(8)=\frac{\pi^8}{9450}$}\hbox{\strut  $\zeta(10)=\frac{\pi^{10}}{93555}$\quad $\zeta(12)=\frac{691\pi^{12}}{638512875}$}} }\hbox{\strut  $\zeta(14)=\frac{2 \pi^{14}}{18243225}$\quad$\zeta(16)=\frac{3617 \pi^{16}}{325641566250}$}} }\hbox{\strut   $\eta(4)=\frac{7\pi^4}{720}$\quad $\eta(6)=\frac{31\pi^6}{31240}$\quad $\eta(8)=\frac{127\pi^8}{1209600}$}} }\hbox{\strut  $\eta(10)=\frac{73\pi^{10}}{6842880}$\quad $\eta(12)=\frac{1414477\pi^{12}}{1307674368000}$}} }\hbox{\strut   $\eta(14)=\frac{8191 \pi^{14}}{74724249600}$\quad$\eta(16)=\frac{16931177 \pi^{16}}{1524374691840000}$}} }\hbox{\strut  $\lambda(4)=\frac{\pi^4}{96}$\quad $\lambda(6)=\frac{\pi^6}{960}$\quad $\lambda(8)=\frac{17\pi^8}{161280}$}} }\hbox{\strut  $\lambda(10)=\frac{31\pi^{10}}{2903040}$\quad $\lambda(12)=\frac{691\pi^{12}}{638668800}$}} }\hbox{\strut  $\lambda(14)=\frac{5461 \pi^{14}}{49816166400}$\quad$\lambda(16)=\frac{929569 \pi^{16}}{83691159552000}$}}  \\                      [1ex] \hline

$\ldots$ &$\ldots$&$\ldots$&$\ldots$\\                      [1ex] 

 \hline\hline 
\end{tabular}

\begin{tablenotes}
\item[a] Sums that appear in the Table are denoted by $\zeta(p)=\sum^{\infty}_{n=1}\frac{1}{n^p}$, $\eta(p)=\sum^{\infty}_{n=1}\frac{(-1)^{n-1}}{n^p}$, $\lambda(p)=\frac{1}{2}(\zeta(p)+\eta(p))=\sum^{\infty}_{n=0}\frac{1}{(2n+1)^p}$.
\item[b] $C$ presents corresponding normalization constant.
    \item[c] $Q_n(x)$ presents $n$th degree polynomial $Q_n(x)=a_nx^n+a_{n-1}x^{n-1}+...+a_1x+a_0$. 
      \end{tablenotes}
\end{threeparttable}
\label{tabelasume}
\end{table}
\end{center}
\end{widetext}

 One can proceed to higher-degree polynomials and obtain analytical forms of sums from the Table \ref{tabelasume} for larger $p$. As a consequence for the higher-degree polynomials and larger value of $p$ the coprime integers appearing in the sums increase. It can be noted that all the sums have irrational values and as $p$ increases the values of sums become closer to $1$. Following the rule of appearance of the sums from Table \ref{tabelasume}, one can conclude that for even polynomial wave functions of degree $n$ sums for $p=2n, p=2,3,...,n$ can be calculated, whereas the odd-degree ones produce the sums for $p=2n-2, p=3,4,...,n$. 

\begin{figure}[ht] 
\includegraphics[width=8.4cm]{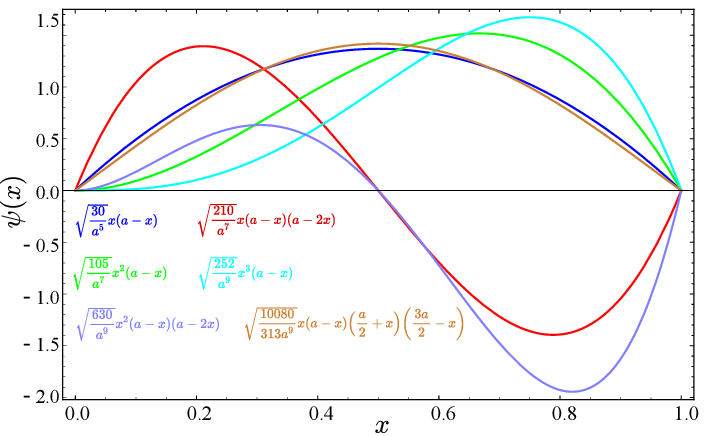}
\centering
\caption{\label{Fig1}
(Color online) Polynomial wave functions discussed in the paper for $a=1$.}
\end{figure}

All polynomial wave functions discussed in the paper are presented in Fig \ref{Fig1}. It can be observed that states $\sqrt{\frac{30}{a^5}}x(a-x)$ and $\sqrt{\frac{10080}{313a^9}}x(a-x)(\frac{a}{2}+x)(\frac{3a}{2}-x)$ are even with respect to the center of the well, whereas the others are not of a specific parity. 
One can check that by increasing the number of nodes of polynomial wave function inside the well (increasing the degree of polynomial), mean energy in the given state increases. Yet, since the wave functions are polynomial only inside the well, it may happen that the higher-degree polynomial wave function has less nodes than the lower-degree one. For instance, the fourth-degree function $\psi(x)=\sqrt{\frac{10080}{313a^9}}x(a-x)(\frac{a}{2}+x)(\frac{3a}{2}-x)$ has no nodes inside the well (two roots are $x=-\frac{a}{2}$ and $x=\frac{3a}{2}$), wherefore the mean energy for the third-degree polynomials with one node (as $\psi(x)=\sqrt{\frac{210}{a^7}}x(a-x)(a-2x)$) is larger.

\section{Conclusion} \label{concl}
We have shown that starting from certain principles of quantum mechanics different infinite sums can be determined analytically. Namely, in the problem of a particle in a box, which is used to illustrate the idea of the method, the calculation of mean value of energy in the polynomial type of state inside the well, leads to different Riemann zeta $\zeta(p)$ and related functions for positive even arguments $p>2$ (similar method that includes the operator $\hat{H}^2$ can be used for computing these sums for p=2; see Appendix \ref{appa}). These sums for odd positive $p$ cannot be calculated analytically by using this or any other method.
An advantage of quantum mechanical approach is that one can check the course of the calculation by employing dimensional analysis based on the constants that appear in the problem ($a$, $\hbar$, $m$). Application of this method to more complex quantum mechanical problems would lead to a variety of other analytically computable infinite sums.  


\appendix
\section{Sums for $p=2$} \label{appa}
Repeating the above performed procedure for the operator $\hat{H}^2$ we can also obtain $\zeta(2)$, $\eta(2)$, and $\lambda(2)$. Following paper \cite{ermitovost}, Eqs. (\ref{srednjajedan}) and (\ref{srednjadva}) are replaced with
\begin{equation} (\hat{H}\psi,\hat{H}\psi)=\sum^{\infty}_{n=1}W(E_n)E_n^2, \end{equation}
where we choose $\psi(x)=\sqrt{\frac{30}{a^5}}x(a-x)$. It was already shown that 
\begin{equation}
W(E_n)=\frac{480}{n^6\pi^6}[1-(-1)^n],
\end{equation}
therefore using
$E_n^2=\frac{n^2\pi^4\hbar^4}{4m^2a^4}$, as well as
$(\hat{H}\psi,\hat{H}\psi)=\frac{30\hbar^4}{m^2a^4}$, we obtain
\begin{equation}
\sum^{\infty}_{n=1}\frac{4}{n^2\pi^2}\left[1-(-1)^n\right]=1. \label{jedan}
\end{equation}
From (\ref{jedan}) and employing (\ref{tajnaveza}) and (\ref{javnaveza}) we get $\zeta(2)=\frac{\pi^2}{6}$, $\eta(2)=\frac{\pi^2}{12}$, and $\lambda(2)=\frac{\pi^2}{8}$. These three sums could be also calculated starting from the polynomial wave function of an arbitrary degree at the expense of solving system of equations.


\end{document}